\documentclass[%
 reprint,
 amsmath,amssymb,
 aps,
 ]{revtex4-2}

\usepackage{graphicx}
\usepackage{dcolumn}
\usepackage{bm}
\usepackage[pdfstartview=FitH,
CJKbookmarks=true,
bookmarksnumbered=true,
bookmarksopen=true,
colorlinks,
linkcolor=blue,
anchorcolor=blue,
citecolor=blue,
allcolors=blue,
pdfborder=001,
]{hyperref}
\usepackage[caption=false]{subfig}
\usepackage{floatrow}
\usepackage{enumitem}
\usepackage{braket}
\DeclareSubrefFormat{myparens}{#1~(#2)}
\DeclareCaptionListOfFormat{myparens}{#1(#2)}
\usepackage{xspace}
\usepackage{multirow}
\usepackage{makecell}
\usepackage{placeins}
\usepackage{subcaption}
\usepackage{ulem}

\begin{document}

\newcommand\MeV{\ensuremath{\mathrm{MeV}}}
\newcommand\GeV{\ensuremath{\mathrm{GeV}}}

\setlength{\abovedisplayskip}{6pt}
\setlength{\belowdisplayskip}{6pt}

\title{\boldmath On the Nature of $X(2370)$}

\author{Di Ben}
\email{bd@mail.tsinghua.edu.cn}
\affiliation{Department of Physics and Center for High Energy Physics, Tsinghua University, Beijing 100084, China}

\author{Li-Ke Yang}
\email{yanglike@itp.ac.cn}
\affiliation{CAS Key Laboratory of Theoretical Physics, Institute of Theoretical Physics, Chinese Academy of
Sciences, Beijing 100190, China\\
School of Physics, University of Chinese Academy of Sciences (UCAS),
Beijing 100049, China}

\author{Bing-Song Zou}
\email{zoubs@mail.tsinghua.edu.cn}
\affiliation{Department of Physics and Center for High Energy Physics, Tsinghua University, Beijing 100084, China}

\date{\today} 

{
\renewcommand{\baselinestretch}{1.20}

\begin{abstract} 
Recently, the BESIII Collaboration claimed that their observed $X(2370)$ ($J^{PC}=0^{-+}$) is predominantly a glueball state based on several arguments. However, we find these arguments to be flawed and the conclusion unjustified. Instead, we demonstrate that all observed properties of the $X(2370)$ can be naturally explained by a $\bar\Sigma\Sigma$ molecule dominated state.
\end{abstract}

\maketitle

A pseudoscalar state, $\eta(2320)$, was observed in $\bar pp$ annihilation into $\eta\eta\eta$ and $\pi\pi\eta$ in the late 1990s and early 2000s~\cite{Zou:1998nr,Anisovich:2000ix}. Since 2011, the BESIII Collaboration has observed the $X(2370)$ in several radiative $J/\psi$ decay channels ($\pi\pi\eta'$, $\pi\pi\eta$, $K\bar K\eta'$ and $K\bar K\pi$), with fitted masses ranging from about 2300 to 2400 MeV~\cite{BESIII:2016fbr,BESIII:2026rzt}.  Its quantum numbers were determined to be $J^{PC}=0^{-+}$ in 2024~\cite{BESIII:2023wfi}. These observations suggest that the $X(2370)$ and the earlier $\eta(2320)$ correspond to the same pseudoscalar state. Thus, the existence of a pseudoscalar state in this mass region is well established. 

More recently, the BESIII Collaboration reported no significant evidence for $X(2370)\to K^*\bar K$ and claimed that the $X(2370)$ is predominantly a glueball state~\cite{BESIII:2026mvn}. The new argument for this claim is that ``the $K^*\bar K$  mode suppression is an unambiguous signature for a $0^{-+}$ flavor-singlet, since a $0^{-+}$ flavor-singlet meson is forbidden to decay into $K^*\bar K$ mode due to the generalized G-parity conservation~\cite{Lipkin:1981uc,Lipkin:1981ak,Klempt:2007cp}", while the flavor-singlet is an important property of glueballs. But there is obviously a logic flaw in this argument: even if a $0^{-+}$ flavor-singlet state is forbidden from decaying into $K^*\bar K$, suppression of the $K^*\bar K$ mode does not by itself imply that a generic $0^{-+}$ state is a flavor singlet. Therefore, the statement in the abstract of Ref.~\cite{BESIII:2026mvn} that it is “the first flavor-singlet light hadron observed above 1 GeV” is not justified.

In fact, the two observed structures, $X(2370)$ and $X(2600)$, lie just below the thresholds of $\bar\Sigma\Sigma$ and $\bar\Xi\Xi$, respectively~\cite{BESIII:2016fbr}.  While the $\eta(2225)$, located just below the $\bar\Lambda\Lambda$ threshold, has been interpreted as a $\bar\Lambda\Lambda$ molecule~\cite{Zhao:2013ffn}, the $\eta(2320)/X(2370)$ and $X(2600)$ are natural candidates for $\bar\Sigma\Sigma$ and $\bar\Xi\Xi$ molecular states, respectively. Such states have been anticipated as light-flavor counterparts of the hidden-charm baryon-antibaryon molecules predicted in Refs.\cite{Dong:2021bvy,Dong:2021juy}. 
The existence of such hyperon-antihyperon molecular states is also supported by QCD sum rule calculations~\cite{Wan:2021vny,Zhang:2025qmg}.

All observed decay modes ($\eta\eta\eta$, $\pi\pi\eta$, $\pi\pi\eta'$, $K\bar K\eta'$ and $K\bar K\pi$) of $\eta(2320)/X(2370)$ share the feature that their quark content can accommodate an $\bar ss\bar q^2 q^2$ configuration with the constituent quarks all in S-wave configurations. This provides further support for a sizable $\bar\Sigma\Sigma$ molecular component, since a $\bar\Sigma\Sigma$ configuration naturally contains the same $\bar ss\bar q^2 q^2$ quark content. The $\eta(2320)/X(2370)$ needs not be a pure $\bar\Sigma\Sigma$ molecule; in general, its wave function may also contain other quark and gluon configurations.
A general Fock-space decomposition of the $\eta(2320)/X(2370)$ wave function may be written schematically as:
\begin{widetext}
\begin{equation}    
|\eta(2320)/X(2370)\rangle = c_{gg}|gg\rangle + c_{\bar qq}|\bar qq\rangle + c_{g\bar qq}|g\bar qq\rangle + c_{\bar q^2q^2}|{\bar q^2q^2}\rangle + c_{\bar q^3q^3}|{\bar q^3q^3}\rangle + \cdots,
\label{Eq1}
\end{equation}
\end{widetext}
where $g$ denotes a gluon and $q$ denotes a light quark (u, d, or s). The quantity $|c_i|^2$ characterizes the weight of the corresponding component. A resonance produced in $J/\psi$ radiative decay evolves from two gluons, as depicted in Fig.~\ref{fig1}, which shows the generic evolution of two gluons into multi-quark states. For two gluons with energy below 1 GeV, they mainly evolve into $\bar qq$ states with multi-quark components much suppressed due to limited energy, such as $\eta$ and $\eta^\prime$ for the case of $0^{-+}$ states. As energy of two gluons increases above 1 GeV, the multi-quark components become more and more important. Around 1.4 GeV, $2^1S_0$ $\bar ss$ state and $a_0\pi$-$\bar K^*K$ dynamically generated states are expected, consistent with the prominent production of $\eta(1410)$ and $\eta(1470)$. Around 1.9 GeV, $3^1S_0$ $\bar ss$ state and $\bar NN$-$\bar K^*_0(1430)K$-$\bar K_1(1270)K^*$-$\bar\Lambda\Lambda$ dynamically generated states are expected. The copiously produced $X(1835)$ and $X(2120)$ may be mixture of these components. Above about 2.3 GeV, six-quark configurations may become increasingly important, naturally accommodating states such as $\eta(2320)/X(2370)$ and $X(2600)$ with large $\bar\Sigma\Sigma$ and $\bar\Xi\Xi$ molecular components, mixed with the $\bar K_1(1400)K^*$ molecule as partner $\bar K_1(1400)K$ molecule~\cite{Dong:2022cuw} and other possible hidden-strangeness tetraquark configurations~\cite{Jafarzade:2025txm}. 

\begin{figure}[H]
    \centering
    {\vglue 0.01cm}
    
    \subfloat[~$gg\to \bar{q}q\to \bar q^2q^2 \to \bar q^3q^3$. ]{
    \includegraphics[width=0.7\columnwidth]{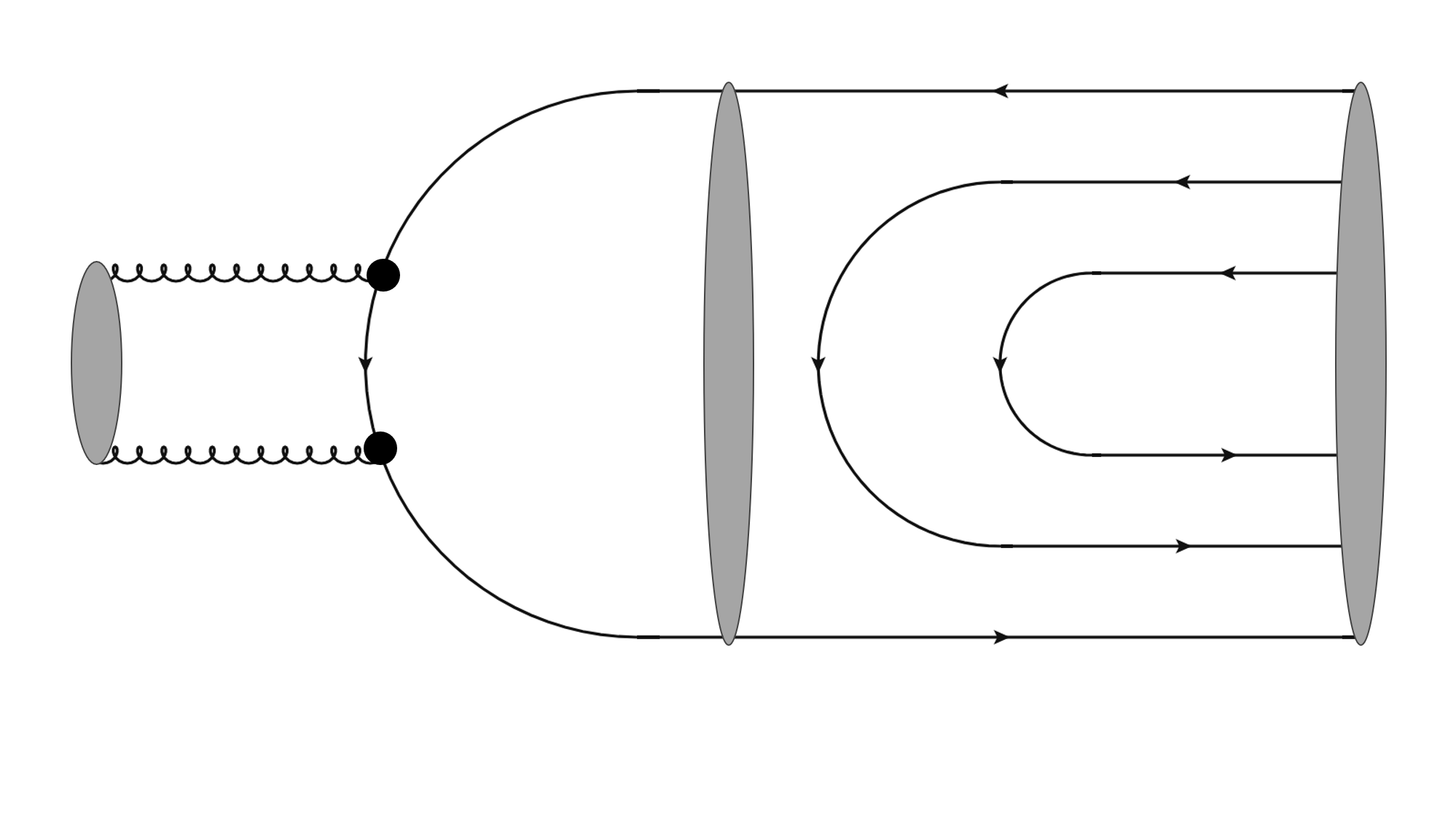}} \\[1pt]
    \subfloat[~$gg\to g\bar{q}q\to \bar q^2q^2 \to \bar q^3q^3$. ]{
    \includegraphics[width=0.7\columnwidth]{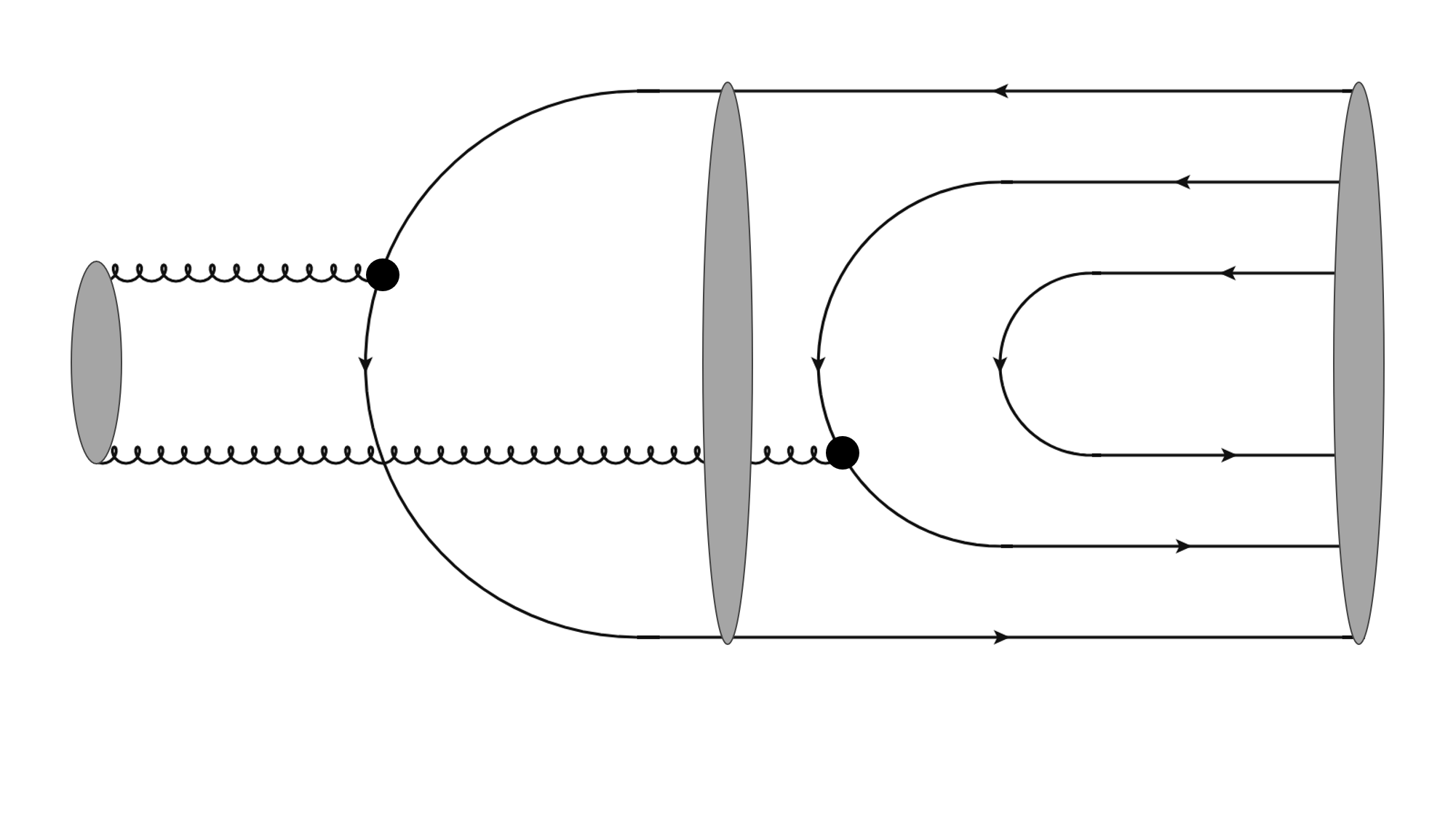}} {\hglue 0.01cm}
    
    \caption{
    Schematic illustration for generic evolution from two gluons to the multiquark states.}
    \label{fig1}
\end{figure}

When a resonance lies close to a hadron-hadron threshold and shares the same quantum numbers as the corresponding S-wave two-hadron state, it is generally expected to contain a sizable hadronic molecular component~\cite{Guo:2017jvc}. The $\eta(2320)/X(2370)$ is therefore expected to contain a sizable $\bar\Sigma\Sigma$ molecular component, providing a natural explanation for its observed decay pattern. However, it may also contain a sizable $\bar ss\bar qq$ component. An analogous situation has been discussed for the $D_{s1}(2460)$,  which, despite its substantial $D^*K$ molecular component, was found to contain about a $54\%$ $c\bar s$ component~\cite{Yang:2021tvc,Hao:2022vwt}. Since the $\bar K^*K$ is a $\bar ss\bar qq$ channel at quark level, the $\eta(2320)/X(2370)$ couples to $\bar K^*K$ through its $\bar ss\bar qq$ component. If the mass difference among $u$, $d$ and $s$ quarks are neglected, the mixed $\bar ss\bar qq$ component has isospin $0$, $U$-spin $0$ and $V$-spin $0$, and therefore does not couple to the $\bar K^*K$ final state just like isospin $0$ pseudo-scalar meson cannot decay to $\rho\pi$.  Hence, the observed suppression of the $K^*\bar K$  mode does not rule out a $\bar\Sigma\Sigma$ molecular interpretation of the $\eta(2320)/X(2370)$.

Another argument used to support the glueball claim for $X(2370)$ is the suppression of radiative decays to $\omega$ and $\phi$~\cite{BESIII:2026mvn}. But for a $\bar\Sigma\Sigma$ molecule of $\bar ss\bar q^2 q^2$ to decay to either $\gamma\omega$ or $\gamma\phi$, it would need a $\bar qq$ or the $\bar ss$ to form the vector meson while the remaining tetra-quarks annihilate to a real photon. Such a process requires substantial quark rearrangement and annihilation and is therefore expected to be strongly suppressed. Consequently, the suppression of the $\gamma\omega$ and $\gamma\phi$ modes does not by itself rule out the $\bar\Sigma\Sigma$ molecular interpretation. 

The third argument presented in Ref.~\cite{BESIII:2026mvn} for claiming the $X(2370)$ as a glueball is the compatibility of the $X(2370)$ mass and quantum numbers with lattice-QCD predictions. However, it is well known that quenched lattice QCD does not represent QCD in the real world, while spectrum calculations in unquenched lattice QCD do not, by themselves, determine the relative weights of the components in Eq.(\ref{Eq1}).  In fact, the $0^{-+}$ state predicted by unquenched lattice QCD could instead have a large $\bar\Sigma\Sigma$ component. The observed mass and spin-parity of $X(2370)$ are in excellent agreement with the expectations for a $\bar\Sigma\Sigma$ molecule.

The fourth argument presented in Ref.~\cite{BESIII:2026mvn} for claiming the $X(2370)$ to be a glueball is its relatively large production rate in $J/\psi$ radiative decays. It is true that the $X(2370)$ is the only observed $0^{-+}$ state above 2.3 GeV with high production branching fraction about $1\times 10^{-3}$ in $J/\psi$ radiative decays. But its production rate in $J/\psi$ radiative decays is much smaller than $\eta^\prime$ and $\eta(1405/1475)$ while comparable with $X(1835)$. Such a production rate is not unexpected for a state with a sizable $\bar\Sigma\Sigma$ molecular component. 

Table~\ref{eta_c} lists the branching fractions of the lowest pseudoscalar charmonium state, $\eta_c$, into stable baryon-antibaryon pairs, as compiled by the PDG~\cite{ParticleDataGroup:2026aaa}.

\begin{table}[htb]
    \centering
    \begin{tabular}{c|c|c}
        $\Gamma$ & Fraction $\Gamma_i/\Gamma$($10^{-3}$) & P(MeV/c) \\\hline
        $\Gamma_{p\bar{p}}$ & $1.11 \pm 0.12$ & 1160\\
        $\Gamma_{\Lambda\bar{\Lambda}}$ & $0.94 \pm 0.18$ & 991\\
        $\Gamma_{\Sigma^+\bar{\Sigma}^-}$ & $1.98 \pm 0.35$ & 901\\
        $\Gamma_{\Xi^-\bar{\Xi}^+}$ & $0.83 \pm 0.18$ & 692
    \end{tabular}
    \caption{Branching fractions of $\eta_c$ into stable baryon-antibaryon pairs and the corresponding relative momenta P. }
    \label{eta_c}
\end{table}
Dividing the branching fractions by the corresponding phase-space factors, which are proportional to P, we obtain the following relation among the $\eta_c$ couplings to baryon-antibaryon channels: 
\begin{widetext}
\begin{equation}
g^2_{\eta_c\Sigma^+\bar{\Sigma}^-}:g^2_{\eta_c\Xi^-\bar{\Xi}^+}:g^2_{\eta_c\Lambda\bar{\Lambda}}:g^2_{\eta_c p\bar{p}} =(2.30\pm 0.48):(1.25\pm 0.30):(0.99\pm 0.22):1
\end{equation}
\end{widetext}
This indicates that $\eta_c$ couples more strongly to the $\Sigma\bar{\Sigma}$ channel than to the $\Xi\bar{\Xi}$, $\Lambda\bar{\Lambda}$ and $N\bar{N}$ channels. This provides further support for the possibility that the nearby $X(2370)$ contains a sizable $\Sigma\bar{\Sigma}$ molecular component. The observed decay width and decay pattern are also compatible with this interpretation.

In addition to the flavor-singlet argument discussed above, Ref.\cite{BESIII:2026mvn} raises two further points against the $\Sigma\bar{\Sigma}$ molecular interpretation: (i) the apparent mismatch between the relatively large production rate of the $X(2370)$ and the smaller production rates of baryon-antibaryon pairs in $J/\psi$ radiative decays, and (ii) the absence of an observable distortion of the $X(2370)$ line shape near the $\Sigma\bar{\Sigma}$ threshold. We address these two issues below.

The BESIII Collaboration fitted the $X(2370)$ peak using a simple constant-width Breit–Wigner parametrization:
\begin{equation}
    BW(s) = \frac{1}{M_X^2-s-iM_X\Gamma_X}.
\label{BW}
\end{equation}
Their latest combined fit to all observed channels~\cite{BESIII:2026rzt} yields 
$M_X = 2359^{+13}_{-14} ~\text{MeV}$ and $\Gamma_X = 170^{+44}_{-29} ~ \text{MeV}$.

If $X(2370)$ has a large $\Sigma\bar{\Sigma}$ molecule component, it should have a large coupling to the nearby $\Sigma\bar{\Sigma}$ threshold. A more appropriate description near the $\Sigma\bar{\Sigma}$ threshold is provided by a Flatté-type parametrization~\cite{Flatte:1976xu}: 
\begin{equation}
    FL(s) = \frac{1}{M_0^2-s-i\sqrt{s}(\Gamma_0+\Gamma_{\Sigma\bar{\Sigma}}(s))}\label{flatte},
\end{equation}
where $\Gamma_0$ represents the contribution from all other decay channels whose thresholds lie far below $M_0$, and $\Gamma_{\Sigma\bar{\Sigma}}(s)$ is the energy dependent partial width for decay into the $\Sigma\bar{\Sigma}$ channel. Note that $M_X$ and $\Gamma_X$ denote the effective Breit–Wigner mass and width extracted by BESIII, whereas $M_0$ and $\Gamma_0$ are the parameters entering the Flatté form.

An effective interaction Lagrangian describing the coupling of the $X(2370)$, with $J^{PC} = 0^{-+}$, to the $\Sigma\bar{\Sigma}$ channel may be written as:
\begin{equation}
    \mathcal{L}_{X\Sigma\Sigma}=g_{X\Sigma\Sigma} \bar{\Sigma}\gamma_5 \Sigma X\label{eq_L},
\end{equation}
where $g_{X\Sigma\Sigma}$ may be estimated using the Weinberg compositeness relation~\cite{Weinberg:1962hj}:
\begin{equation}
    \frac{g_{X\Sigma\Sigma}^2}{4\pi} = 4(1-Z)\sqrt{\frac{B_E}{m_\Sigma}}.
\end{equation}
Here Z denotes the fraction associated with non-$\Sigma\bar{\Sigma}$ components, and $B_E$ is the $\Sigma\bar{\Sigma}$ binding energy.
Then the $\Gamma_{\Sigma\bar{\Sigma}}(s)$ can be obtained as:
\begin{equation}
    \Gamma_{\Sigma\bar{\Sigma}}(s) = \frac{g_{X\Sigma\Sigma}^2}{4\pi}\sqrt{\frac{s}{4}-m^2_\Sigma}.
\end{equation}

Assuming a compositeness of $(1-Z) = 0.5$ and choosing $M_0$ and $\Gamma_0$ to approximately reproduce the observed $M_X$ and width $\Gamma_X$, we obtain $M_0 = 2330$ MeV and $\Gamma_0 = 310$ MeV. In Fig.~\ref{FL}, the amplitudes squared for the two kinds of parametrization are shown for comparison. The blue solid line represents the Breit–Wigner line shape using Eq.(\ref{BW}), the red dashed curve corresponds to the Flatté parameterization in Eq.(\ref{flatte}), and the gray vertical line indicates the $\Sigma\bar{\Sigma}$ threshold. The sizable coupling to the $\Sigma\bar{\Sigma}$ channel produces a visible distortion of the line shape relative to the simplified constant-width Breit–Wigner parametrization, which neglects the energy dependence of the partial widths and effective couplings. Given the sizable uncertainties in the masses and widths extracted from the BESIII fits to the various channels, we do not expect a distortion of this magnitude to be clearly resolvable with the current data. 

\begin{figure}[H]
    \centering
    \includegraphics[width=0.6\textwidth]{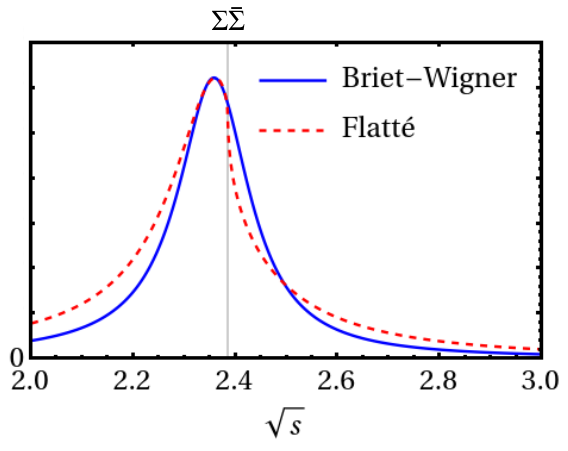}
    \caption{Comparison of the line shapes with the Breit-Wigner formula (blue solid line) and Flatté parameterization (red dashed line).}
    \label{FL}
\end{figure}

Based on the above assumption and the Flatté parameterization of the line shape, we further compute the branching fraction for the decay chain $J/\psi \to \gamma X \to \gamma \Sigma \bar{\Sigma}$ using the Feynman diagrams shown in Fig.~\ref{Ratio}.

\begin{figure}[H]

   \centering
    {\vglue 0.01cm}
    \subfloat[~$J/\psi \to \gamma X \to \gamma \Sigma \bar{\Sigma}$.]{
    \includegraphics[width=0.7\columnwidth]{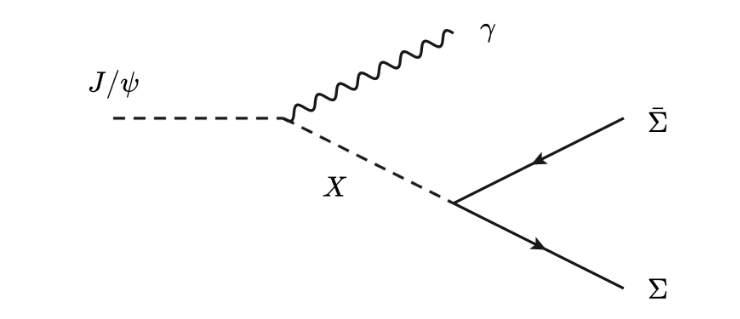}} \\[1pt]
    \vglue 6pt
    \subfloat[~$J/\psi \to \gamma X$ ]{
    \includegraphics[width=0.7\columnwidth]{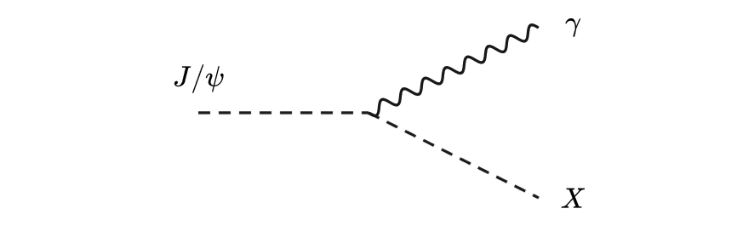}} {\hglue 0.01cm}
    \caption{Feynman diagrams contributing to the ratio defined in Eq.(\ref{R}).}
    \label{Ratio}
\end{figure}

The effective interaction for $J/\psi \to \gamma X$ may be written as:
\begin{equation}
    \mathcal{L}_{\gamma X\psi} = g_{\gamma X\psi} \epsilon^{\alpha\mu\lambda\nu}(\partial_\alpha A_\mu)(\partial_\lambda X)\psi_\nu,
\end{equation}
where $A_\mu$ denoting  electromagnetic field. The specific value of the coupling constant $g_{\gamma X\psi}$ is irrelevant, as it cancels out in the calculation.
We then obtain:
\begin{equation}
    R=\frac{\Gamma_{J/\psi\to\gamma X\to\gamma\Sigma\bar{\Sigma}}}{\Gamma_{J/\psi\to\gamma X}}\sim7\% .
\label{R}
\end{equation}
Note that in the Flatté parameterization for the $X(2370)$ its mass and decay width are $M_0 = 2330$ MeV and $\Gamma_0 = 310$ MeV. Its mass is about 50 MeV below the $\Sigma\bar{\Sigma}$ threshold. It has a large total decay width to various channels below the $\Sigma\bar{\Sigma}$ threshold. Its peak width looks much narrower than its decay width due to its strong coupling to the
$\Sigma\bar{\Sigma}$, just like the case for the $f_0(980)$ where the strong coupling to the $K\bar K$  makes its peak width much smaller than its decay width~\cite{Zou:1993az}. For the $J/\psi\to\gamma X\to\gamma\Sigma\bar{\Sigma}$, the $X(2370)$ decays to $\Sigma\bar{\Sigma}$ from its threshold to the mass of $J/\psi$ with suppression of $\Sigma\bar{\Sigma}$ phase space factor at lower energy end and $q^3_\gamma$ suppression of $J/\psi\to\gamma X$ at higher energy end. This makes the value to be small. In fact, in the Flatté parameterization of Eq.(\ref{flatte}) the decay width ignores its couplings to the channels with thresholds above the $\Sigma\bar{\Sigma}$ threshold which could further suppressing the tail of the $X(2370)$. To take into account such effects an empiric off-shell form factor is usually used~\cite{Liang:2004sd}. If an off-shell form factor of a commonly used form $F = 1/[1+(s-m_0^2)^2/\Lambda^4]$ is included in our calculation with $\Lambda$ around 1~GeV the branching ratio will further suppressed to be $R\sim 3\%$.
Taking $B(J/\psi \to \gamma X) \sim 10^{-3}$, we estimate $B(J/\psi \to \gamma X\to\gamma \Sigma\bar{\Sigma})$ to be around $3 \times 10^{-5}$.

Therefore, the relatively large production rate of the $X(2370)$ and the small baryon-antibaryon production rates in radiative $J/\psi$ decays can be simultaneously accommodated within the $\Sigma\bar{\Sigma}$ molecular interpretation.

Last but not least, two-photon production provides an important complementary probe to radiative $J/\psi$ decays in searches for glueball candidates. Whereas $J/\psi\to\gamma X$ provides a gluon-rich production environment, two-photon production is more sensitive to states with substantial electrically charged quark content. A state that is strongly produced in radiative $J/\psi$ decays but only weakly produced in two-photon collisions would therefore be expected to have a comparatively large gluonic component. Conversely, strong two-photon production would point toward a substantial quark component.
Two-photon production of $\eta^\prime\pi^+\pi^-$ was studied by Belle Collaboration~\cite{Belle:2012uhr}. Below 2.9 GeV, the X(2370) appears as the only prominent structure in Fig.2(b) of Ref.~\cite{Belle:2012uhr}. In the same $\eta^\prime\pi^+\pi^-$ final state produced in radiative $J/\psi$ decay, four structures associated with $X(1835)$, $X(2120)$, $X(2370)$, and $X(2600)$ are observed, with $X(1835)$ being the most prominent~\cite{BESIII:2016fbr}. This observation further favors an interpretation of the$ X(2370)$ as a state with a sizable $\Sigma\bar{\Sigma}$ molecular component over a predominantly glueball interpretation. 

In conclusion, the $X(2370)$ observed by the BESIII Collaboration is consistent with the earlier $\eta(2320)$~\cite{Zou:1998nr,Anisovich:2000ix}. We find that all the key arguments in Ref.~\cite{BESIII:2026mvn} for claiming it as a glueball-dominated state are flawed. Rather, a $\Sigma\bar{\Sigma}$ molecule dominated state provides a satisfactory description of all the observed properties of the $X(2370)$.

Based on our conclusion, we predict that (i) the $X(2370)$ can be found in $J/\psi\to\gamma\eta\eta\eta$ with $X(2370)\to f_0(1500)\eta\to\eta\eta\eta$ as its prominent contribution; (ii) the $X(2600)$ has $J^{PC}=0^{-+}$, which can be measured from its decay to $\Sigma\bar\Sigma$ or $\Lambda\bar\Lambda$.  

\smallskip
We thank Feng-Kun Guo, Jia-Jun Wu, Zhi-Hong Ye, Ke Wang, Mao-Jun Yan and Xiang-Yang Xu for useful discussions.

\bibliographystyle{apsrev4-2}
\bibliography{ref}

@article{Zou:1993az,
    author = "Zou, B. S. and Bugg, D. V.",
    title = "{Is f0 (975) a narrow resonance?}",
    doi = "10.1103/PhysRevD.48.R3948",
    journal = "Phys. Rev. D",
    volume = "48",
    pages = "R3948--R3952",
    year = "1993"
}

@article{Liang:2004sd,
    author = "Liang, Wei-Hong and Shen, Peng-Nian and Zou, Bing-Song and Faessler, Amand",
    title = "{Nucleon pole contributions in J / psi ---{\ensuremath{>}} N anti-N pi, p anti-p eta, p anti-p eta-prime and p anti-p omega decays}",
    eprint = "nucl-th/0404024",
    archivePrefix = "arXiv",
    doi = "10.1140/epja/i2004-10007-y",
    journal = "Eur. Phys. J. A",
    volume = "21",
    pages = "487--500",
    year = "2004"
}

@article{Jafarzade:2025txm,
    author = "Jafarzade, Shahriyar and Lebed, Richard F.",
    title = "{Diabatic dynamical diquark model of hidden-strangeness tetraquarks}",
    eprint = "2510.15844",
    archivePrefix = "arXiv",
    primaryClass = "hep-ph",
    doi = "10.1103/vspf-jybk",
    journal = "Phys. Rev. D",
    volume = "112",
    number = "11",
    pages = "114047",
    year = "2025"
}

@article{Zou:1998nr,
    author = "Zou, Bing-Song",
    editor = "Cicalo, C. and De Falco, A. and Puddu, Giovanna and Serci, S.",
    title = "{Search for glueballs from three-body annihilation of anti-p p inflight}",
    eprint = "hep-ex/9812007",
    archivePrefix = "arXiv",
    doi = "10.1016/S0375-9474(99)00178-5",
    journal = "Nucl. Phys. A",
    volume = "655",
    pages = "41--50",
    year = "1999"
}

@article{BESIII:2026rzt,
    author = "Ablikim, Medina and others",
    collaboration = "BESIII",
    title = "{Observation of the $X(2370)$ in $J/ψ\rightarrowγK^{0}_{S}K^{0}_{S}π^{0}$ and $J/ψ\rightarrowγπ^{0}π^{0}η$}",
    eprint = "2605.26495",
    archivePrefix = "arXiv",
    primaryClass = "hep-ex",
    journal = "",
    month = "5",
    year = "2026"
}

@article{Dong:2022cuw,
    author = "Dong, Xiang-Kun and Lin, Yong-Hui and Zou, Bing-Song",
    title = "{Interpretation of the {\ensuremath{\eta}}$_{1}$ (1855) as a KK̄$_{1}$(1400) + c.c. molecule}",
    eprint = "2202.00863",
    archivePrefix = "arXiv",
    primaryClass = "hep-ph",
    doi = "10.1007/s11433-022-1887-5",
    journal = "Sci. China Phys. Mech. Astron.",
    volume = "65",
    number = "6",
    pages = "261011",
    year = "2022"
}

@article{Guo:2017jvc,
    author = "Guo, Feng-Kun and Hanhart, Christoph and Mei{\ss}ner, Ulf-G. and Wang, Qian and Zhao, Qiang and Zou, Bing-Song",
    title = "{Hadronic molecules}",
    eprint = "1705.00141",
    archivePrefix = "arXiv",
    primaryClass = "hep-ph",
    doi = "10.1103/RevModPhys.90.015004",
    journal = "Rev. Mod. Phys.",
    volume = "90",
    number = "1",
    pages = "015004",
    year = "2018",
    note = "[Erratum: Rev.Mod.Phys. 94, 029901 (2022)]"
}

@article{Hao:2022vwt,
    author = "Hao, Wei and Lu, Yu and Zou, Bing-Song",
    title = "{Coupled channel effects for the charmed-strange mesons}",
    eprint = "2208.10915",
    archivePrefix = "arXiv",
    primaryClass = "hep-ph",
    doi = "10.1103/PhysRevD.106.074014",
    journal = "Phys. Rev. D",
    volume = "106",
    number = "7",
    pages = "074014",
    year = "2022"
}

@article{Yang:2021tvc,
    author = "Yang, Zhi and Wang, Guang-Juan and Wu, Jia-Jun and Oka, Makoto and Zhu, Shi-Lin",
    title = "{Novel Coupled Channel Framework Connecting the Quark Model and Lattice QCD for the Near-threshold Ds States}",
    eprint = "2107.04860",
    archivePrefix = "arXiv",
    primaryClass = "hep-ph",
    doi = "10.1103/PhysRevLett.128.112001",
    journal = "Phys. Rev. Lett.",
    volume = "128",
    number = "11",
    pages = "112001",
    year = "2022"
}

@article{Belle:2012uhr,
    author = "Zhang, C. C. and others",
    collaboration = "Belle",
    title = "{First study of $\eta_c$, $\eta(1760)$ and $X(1835)$ production via $\eta^\prime\pi^+\pi^-$ final states in two-photon collisions}",
    eprint = "1206.5087",
    archivePrefix = "arXiv",
    primaryClass = "hep-ex",
    reportNumber = "BELLE-PREPRINT-2010-17, KEK-PREPRINT-2010-26",
    doi = "10.1103/PhysRevD.86.052002",
    journal = "Phys. Rev. D",
    volume = "86",
    pages = "052002",
    year = "2012"
}

@article{Flatte:1976xu,
    author = "Flatte, Stanley M.",
    title = "{Coupled - Channel Analysis of the pi eta and K anti-K Systems Near K anti-K Threshold}",
    reportNumber = "CERN-EP-PHYS-76-8",
    doi = "10.1016/0370-2693(76)90654-7",
    journal = "Phys. Lett. B",
    volume = "63",
    pages = "224--227",
    year = "1976"
}

@article{BESIII:2016fbr,
    author = "Ablikim, Medina and others",
    collaboration = "BESIII",
    title = "{Observation of an anomalous line shape of the $\eta^{\prime}\pi^{+}\pi^{-}$ mass spectrum near the $p\bar{p}$ mass threshold in $J/\psi\rightarrow\gamma\eta^{\prime}\pi^{+}\pi^{-}$}",
    eprint = "1603.09653",
    archivePrefix = "arXiv",
    primaryClass = "hep-ex",
    doi = "10.1103/PhysRevLett.117.042002",
    journal = "Phys. Rev. Lett.",
    volume = "117",
    number = "4",
    pages = "042002",
    year = "2016"
}

@article{BESIII:2023wfi,
    author = "Ablikim, Medina and others",
    collaboration = "BESIII",
    title = "{Determination of Spin-Parity Quantum Numbers of X(2370) as 0-+ from J/{\ensuremath{\psi}}{\textrightarrow}{\ensuremath{\gamma}}KS0KS0{\ensuremath{\eta}}'}",
    eprint = "2312.05324",
    archivePrefix = "arXiv",
    primaryClass = "hep-ex",
    doi = "10.1103/PhysRevLett.132.181901",
    journal = "Phys. Rev. Lett.",
    volume = "132",
    number = "18",
    pages = "181901",
    year = "2024"
}

@article{BESIII:2026mvn,
    author = "Ablikim, Medina and others",
    collaboration = "BESIII",
    title = "{Lightest $0^{-+}$ Glueball as Dominant Constituent of $X(2370)$}",
    eprint = "2607.20366",
    archivePrefix = "arXiv",
    primaryClass = "hep-ex",
    month = "7",
    journal = "",
    year = "2026"
}

@article{Anisovich:2000ix,
    author = "Anisovich, A. V. and Baker, C. A. and Batty, C. J. and Bugg, D. V. and Nikonov, V. A. and Sarantsev, A. V. and Sarantsev, V. V. and Zou, B. S.",
    title = "{A study of anti-p p --{\ensuremath{>}} eta eta eta for masses 1960-MeV/c**2 to 2410-MeV/c**2}",
    doi = "10.1016/S0370-2693(00)01301-0",
    journal = "Phys. Lett. B",
    volume = "496",
    pages = "145--153",
    year = "2000"
}

@article{Lipkin:1981uc,
    author = "Lipkin, Harry J.",
    title = "{The E Is Not a Glue Ball: But Flavor Symmetry Shows How to Find Them}",
    reportNumber = "FERMILAB-PUB-81-066-T, ANL-HEP-PR-81-23",
    doi = "10.1016/0370-2693(81)91092-3",
    journal = "Phys. Lett. B",
    volume = "106",
    pages = "114--118",
    year = "1981"
}

@article{Lipkin:1981ak,
    author = "Lipkin, Harry J.",
    title = "{GLUEBALLS versus QUARKONIUM: FLAVOR SYMMETRY SIGNATURES}",
    reportNumber = "ANL-HEP-PR-81-35",
    doi = "10.1016/0370-2693(82)90445-2",
    journal = "Phys. Lett. B",
    volume = "109",
    pages = "326--330",
    year = "1982"
}

@article{Klempt:2007cp,
    author = "Klempt, Eberhard and Zaitsev, Alexander",
    title = "{Glueballs, Hybrids, Multiquarks. Experimental facts versus QCD inspired concepts}",
    eprint = "0708.4016",
    archivePrefix = "arXiv",
    primaryClass = "hep-ph",
    doi = "10.1016/j.physrep.2007.07.006",
    journal = "Phys. Rept.",
    volume = "454",
    pages = "1--202",
    year = "2007"
}

@article{Zhao:2013ffn,
    author = "Zhao, Lu and Li, Ning and Zhu, Shi-Lin and Zou, Bing-Song",
    title = "{Meson-exchange model for the $\Lambda\bar{\Lambda}$ interaction}",
    eprint = "1302.1770",
    archivePrefix = "arXiv",
    primaryClass = "hep-ph",
    doi = "10.1103/PhysRevD.87.054034",
    journal = "Phys. Rev. D",
    volume = "87",
    number = "5",
    pages = "054034",
    year = "2013"
}

@article{Dong:2021bvy,
    author = "Dong, Xiang-Kun and Guo, Feng-Kun and Zou, Bing-Song",
    title = "{A survey of heavy{\textendash}heavy hadronic molecules}",
    eprint = "2108.02673",
    archivePrefix = "arXiv",
    primaryClass = "hep-ph",
    doi = "10.1088/1572-9494/ac27a2",
    journal = "Commun. Theor. Phys.",
    volume = "73",
    number = "12",
    pages = "125201",
    year = "2021"
}

@article{Dong:2021juy,
    author = "Dong, Xiang-Kun and Guo, Feng-Kun and Zou, Bing-Song",
    title = "{A survey of heavy-antiheavy hadronic molecules}",
    eprint = "2101.01021",
    archivePrefix = "arXiv",
    primaryClass = "hep-ph",
    doi = "10.13725/j.cnki.pip.2021.02.001",
    journal = "Progr. Phys.",
    volume = "41",
    pages = "65--93",
    year = "2021"
}

@article{ParticleDataGroup:2026aaa,
    author = "Takahashi, F. and others",
    collaboration = "Particle Data Group",
    title = "{Review of Particle Physics}",
    doi = "10.1142/S0217751X26300115",
    journal = "Int. J. Mod. Phys. A",
    volume = "41",
    pages = "2630011",
    year = "2026"
}

@article{Weinberg:1962hj,
    author = "Weinberg, Steven",
    title = "{Elementary particle theory of composite particles}",
    doi = "10.1103/PhysRev.130.776",
    journal = "Phys. Rev.",
    volume = "130",
    pages = "776--783",
    year = "1963"
}

@article{Wan:2021vny,
    author = "Wan, Bing-Dong and Zhang, Sheng-Qi and Qiao, Cong-Feng",
    title = "{Light baryonium spectrum}",
    eprint = "2109.07130",
    archivePrefix = "arXiv",
    primaryClass = "hep-ph",
    doi = "10.1103/PhysRevD.105.014016",
    journal = "Phys. Rev. D",
    volume = "105",
    number = "1",
    pages = "014016",
    year = "2022"
}

@article{Zhang:2025qmg,
    author = "Zhang, Sheng-Qi and Qiao, Cong-Feng",
    title = "{Baryons and baryoniums in the perspective of QCD sum rules}",
    eprint = "2512.24706",
    archivePrefix = "arXiv",
    primaryClass = "hep-ph",
    doi = "10.1007/s43673-026-00192-y",
    journal = "AAPPS Bull.",
    volume = "36",
    number = "1",
    pages = "12",
    year = "2026"
}

\end{document}